\documentstyle[12pt]{article}

\def\fund{  \> {\vcenter  {\vbox
               {\hrule height.6pt
                \hbox {\vrule width.6pt  height5pt
                      \kern5pt
                      \vrule width.6pt  height5pt}
                \hrule height.6pt}
                         }
               }
            \>\>  }

\def\antifund{  \> \overline{ {\vcenter  {\vbox
               {\hrule height.6pt
                \hbox {\vrule width.6pt  height5pt
                      \kern5pt
                      \vrule width.6pt  height5pt}
                \hrule height.6pt}
                         }
               } }
            \>\>  }

\def\sym{  \> {\vcenter  {\vbox
              {\hrule height.6pt
               \hbox {\vrule width.6pt  height5pt
                      \kern5pt
                      \vrule width.6pt  height5pt
                      \kern5pt
                      \vrule width.6pt  height5pt}
               \hrule height.6pt}
                         }
               }
            \>\>  }

\def\symbar{  \> \overline{ {\vcenter  {\vbox
              {\hrule height.6pt
               \hbox {\vrule width.6pt  height5pt
                      \kern5pt
                      \vrule width.6pt  height5pt
                      \kern5pt
                      \vrule width.6pt  height5pt}
               \hrule height.6pt}
                         }
               } }
            \>\>  }

\def\antisym{ \> {\vcenter  {\vbox
                 {\hrule height.6pt
                  \hbox {\vrule width.6pt  height5pt
                         \kern5pt
                         \vrule width.6pt  height5pt}
                  \hrule height.6pt
                  \hbox {\vrule width.6pt  height5pt
                         \kern5pt
                         \vrule width.6pt  height5pt}
               \hrule height.6pt}
                         }
               }
            \>\>  }

\def\antithree{ \> {\vcenter  {\vbox
                 {\hrule height.6pt
                  \hbox {\vrule width.6pt  height5pt
                         \kern5pt
                         \vrule width.6pt  height5pt}
                  \hrule height.6pt
                  \hbox {\vrule width.6pt  height5pt
                         \kern5pt
                         \vrule width.6pt  height5pt}
                  \hrule height.6pt
                  \hbox {\vrule width.6pt  height5pt
                         \kern5pt
                         \vrule width.6pt  height5pt}
               \hrule height.6pt}
                         }
               }
            \>\>  }

\def\antifour{ \> {\vcenter  {\vbox
                 {\hrule height.6pt
                  \hbox {\vrule width.6pt  height5pt
                         \kern5pt
                         \vrule width.6pt  height5pt}
                  \hrule height.6pt
                  \hbox {\vrule width.6pt  height5pt
                         \kern5pt
                         \vrule width.6pt  height5pt}
                  \hrule height.6pt
                  \hbox {\vrule width.6pt  height5pt
                         \kern5pt
                         \vrule width.6pt  height5pt}
                  \hrule height.6pt
                  \hbox {\vrule width.6pt  height5pt
                         \kern5pt
                         \vrule width.6pt  height5pt}
               \hrule height.6pt}
                         }
               }
            \>\>  }

\def\antifive{ \> {\vcenter  {\vbox
                 {\hrule height.6pt
                  \hbox {\vrule width.6pt  height5pt
                         \kern5pt
                         \vrule width.6pt  height5pt}
                  \hrule height.6pt
                  \hbox {\vrule width.6pt  height5pt
                         \kern5pt
                         \vrule width.6pt  height5pt}
                  \hrule height.6pt
                  \hbox {\vrule width.6pt  height5pt
                         \kern5pt
                         \vrule width.6pt  height5pt}
                  \hrule height.6pt
                  \hbox {\vrule width.6pt  height5pt
                         \kern5pt
                         \vrule width.6pt  height5pt}
                  \hrule height.6pt
                  \hbox {\vrule width.6pt  height5pt
                         \kern5pt
                         \vrule width.6pt  height5pt}
               \hrule height.6pt}
                         }
               }
            \>\>  }

\def\antisix{ \> {\vcenter  {\vbox
                 {\hrule height.6pt
                  \hbox {\vrule width.6pt  height5pt
                         \kern5pt
                         \vrule width.6pt  height5pt}
                  \hrule height.6pt
                  \hbox {\vrule width.6pt  height5pt
                         \kern5pt
                         \vrule width.6pt  height5pt}
                  \hrule height.6pt
                  \hbox {\vrule width.6pt  height5pt
                         \kern5pt
                         \vrule width.6pt  height5pt}
                  \hrule height.6pt
                  \hbox {\vrule width.6pt  height5pt
                         \kern5pt
                         \vrule width.6pt  height5pt}
                  \hrule height.6pt
                  \hbox {\vrule width.6pt  height5pt
                         \kern5pt
                         \vrule width.6pt  height5pt}
                  \hrule height.6pt
                  \hbox {\vrule width.6pt  height5pt
                         \kern5pt
                         \vrule width.6pt  height5pt}
               \hrule height.6pt}
                         }
               }
            \>\>  }

\begin{document}
\baselineskip 18pt

\vspace*{5mm}

\begin{flushright}
DPNU-98-01 \\
January 1998 
\end{flushright}

\vspace{1cm}

\begin{center}
{\large\bf Confining phase in SUSY SO(12) gauge theory 
with one spinor and several vectors}
\end{center}

\vspace{1cm}

\begin{center}
{\large Nobuhito Maru} 
\end{center}

\begin{center}
Department of Physics \\ 
Nagoya University \\ 
Nagoya 464-0814, JAPAN 
\end{center}

\begin{center} 
{\normalsize {\tt maru@eken.phys.nagoya-u.ac.jp}} 
\end{center}

\vspace{2cm}
\setcounter{page}{0}
\thispagestyle{empty}
\begin{center}
{\bf Abstract}
\end{center}
We study the confining phase structure of ${\cal N}$=1 
supersymmetric 
$SO(12)$ gauge theory with $N_f \le 7$ vectors 
and one spinor. 
The explicit form of low energy superpotentials 
for $N_f \le 7$ are derived 
after gauge invariant operators relevant in 
the effective theory 
are identified via gauge symmetry breaking pattern. 
The resulting confining phase structure is analogous to 
$N_f \le N_c+1$ SUSY QCD. Finally, we conclude with some comments on 
the search for duals to $N_f \ge 8$ $SO(12)$ theory. 

\vspace{1cm}

\begin{flushleft}
\end{flushleft}

\newpage

\section{Introduction}
Our understanding of non-perturbative nature in ${\cal N}=1$ 
supersymmetric gauge theories has much progressed 
since the pioneering works of Seiberg and his collaborators 
\cite{Seiberg,ILS}. Especially, physics of confining phase in 
${\cal N}=1$ supersymmetric gauge theories was enriched 
(quantum deformed moduli space, ``s-confinement"). 
These works have been also extended to 
the theories with various types of gauge groups 
and matter contents 
\cite{s-conf,so11,GN,exc}. 
Furthermore, these theories have recently been 
applied to construction of models with dynamical supersymmetry breaking 
or SUSY composite models.

In this paper, we study the confining phase 
in ${\cal N}=1$ supersymmetric $SO(12)$ gauge theory with 
$N_f \le 7$ vectors and one spinor. 
There are two motivations we are interested in 
this particular model. 
First, from the theoretical point of view, 
it will provide useful informations for finding 
the dual to $SO(N_c)(N_c > 10)$ with an arbitrary number of 
vectors and spinors. Although the duality of this class of models 
has only been generalized to $SO(10)$ \cite{BCKS}, 
the known dualities have the following remarkable properties 
which are not contained in Seiberg's duality \cite{Seiberg1}: 
1. Chiral-Nonchiral duality. 
2. Reducibility to the exceptional group ($G_2$) duality. 
3. Simple and Semi-simple group duality 
(without ``deconfinement"). 
4. Identification of massive spinors and 
$Z_2$ monopoles under duality 
\cite{BCKS,Pouliot,so8,so10,Kawano,more,Strassler}. 
Therefore, it is natural to ask whether these properties exist 
in the duality for $SO(N_c)(N_c > 10)$ theory. 
However, looking for this dual seems to be highly 
non-trivial from 
the result of Ref. \cite{BCKS}. Cho \cite{so11} has already 
investigated in detail the confining phase of $SO(11)$ 
gauge theory with $N_f \le 6$ vectors and a spinor 
and extracted some clues in search for duals. 
It is interesting enough to pursue further following the line of 
his argument in order to clarify the dual to $SO(N_c)(N_c > 10)$ 
theory. Second, as mentioned in the above paragraph, 
the theory under consideration may provide 
phenomenologically viable models with dynamical 
supersymmetry breaking or SUSY composite models.

This paper is organized as follows. 
In section 2, gauge invariant operators relevant 
to the low energy physics are identified. 
We derive the explicit form of low energy superpotentials 
for $N_f \le 7$ in section 3. In the last section, summary and 
some comments on the search for duals to 
$N_f \ge 8$ $SO(12)$ theory 
are given.

\section{The $SO(12)$ model}
The model we consider has following symmetry groups
\begin{equation}
\label{sym}
G = SO(12)_{gauge} \times [SU(N_f)_V \times U(1)_V \times 
U(1)_Q \times U(1)_R]_{global}
\end{equation}
under which the superfields transform as\footnote{We implicitly 
regard the 32 dimensional $SO(12)$ spinor as the projection 
$Q=P_- Q_{64}$ where $Q_{64}$ means the 64 dimensional spinor of 
$SO(13)$ and $P_- = \frac{1}{2}(1-\Gamma_{13})$.} 
\begin{eqnarray}
\label{matter}
V^i_{\mu} &\sim& ({\bf 12}, \fund, 1, 0, 0), \\
Q_{\alpha} &\sim& ({\bf 32}, {\bf1}, 0, 1, 0)
\end{eqnarray}
and no tree level superpotential. 
Note that since each of the $U(1)$ symmetries in Eq.(\ref{sym}) are 
anomalous, the action is transformed as 
\begin{equation}
S \to S - i C \alpha \int d^4x \frac{g^2}{32 \pi^2} F \tilde{F},
\end{equation}
where $C$ denotes the anomaly coefficient of the corresponding 
$U(1)_VSO(12)^2$, $U(1)_QSO(12)^2$ or $U(1)_RSO(12)^2$ anomalies 
and $\alpha$ is a transformation parameter. 
If the theta parameter in the Lagrangian is shifted 
under these anomalous $U(1)$'s as 
$\theta \to \theta + C \alpha$, then anomalies can be cancelled. 
Recalling the relation
\begin{equation}
\label{1-loop RGE}
\left( \frac{\Lambda}{\mu} \right)^{b_0} = {\rm exp} 
\left( - \frac{8\pi^2}{g^2(\mu)} + i \theta \right), 
\end{equation}
where $b_0$ represents 1-loop beta function coefficient
\footnote{$\mu$ denotes quadratic Dynkin index defined as 
Tr $T^a(R)T^b(R) = \mu(R) \delta^{ab}$ 
($T^a$: the generators of the group, 
$R$: representation which the superfield belongs to). 
We use the following values : 
$\mu({\bf 12}) = 2$, $\mu({\bf 32}) = 8$, $\mu({\bf 66}) = 20$.}
\begin{equation}
\label{beta}
b_0 = \frac{1}{2} [3 \mu({\rm Adj}) - 
\sum_{{\rm matter}} \mu(R)] = 
26 - N_f, 
\end{equation}
and $\Lambda$ is the strong coupling scale of the theory, 
the spurion superfield $\Lambda^{b_0}$ is transformed as 
\begin{equation}
\label{spurion}
\Lambda^{b_0} \sim (1, 1, 2N_f, 8, 12-2N_f). 
\end{equation}






Using these symmetries and holomorphy, we can easily fix 
the form of the dynamicaly generated superpotential $W_{dyn}$ 
for the small value of $N_f$ so that $U(1)_R$ charge of $W_{dyn}$ 
be $2$ and $U(1)_V, U(1)_Q$ charges vanish. 
The results are summarized in Table 1. 
%
\begin{table}
\begin{center}
    \begin{tabular}{ccc}
\hline
$N_f$ & $R(\Lambda^{b_0})$ & $W_{dyn}$ \\
\hline
0 & 12 & $(\Lambda^{26}/Q^8)^{1/6}$ \\ 
1 & 10 & $(\Lambda^{25}/Q^8 V^2)^{1/5}$ \\ 
2 & 8 & $(\Lambda^{24}/Q^8 V^4)^{1/4}$ \\ 
3 & 6 & $(\Lambda^{23}/Q^8 V^6)^{1/3}$ \\
4 & 4 & $(\Lambda^{22}/Q^8 V^8)^{1/2}$ \\ 
5 & 2 & $\Lambda^{21}/Q^8 V^{10}$ \\ 
6 & 0 & $X(Q^8 V^{12} - \Lambda^{20})$ \\ 
7 & $-2$ & $Q^8 V^{14}/\Lambda^{19}$ \\
\hline
\end{tabular}
\end{center}
\caption[sp]{The form of dynamically generated superpotentials}
\end{table}
%
$N_f=6$ case is special because R charge of $\Lambda^{b_0}$ vanishes, 
therefore we cannot construct the dyamically generated superpotential. 
However, the classical constraint among matter superfields is 
modified by non-perturbative effects and this quantum constraint 
can be included in the superpotential 
by using the Lagrange mutiplier superfield $X$ \cite{Seiberg}. 
Therefore $N_f=6$ case is analogous to $N_f=N_c$ SUSY QCD 
(quantum deformation of moduli space). 
Furthermore, $N_f=7$ case is analogous to $N_f=N_c+1$ SUSY QCD 
(``s-confinement") \cite{Seiberg} 
and $N_f \le 5$ case is analogous to $N_f \le N_c-1$ 
SUSY QCD (runaway superpotential) \cite{ADS}.

In order to describe the low energy effective theory, 
we need to find gauge invariant operators which behave as 
the moduli space coordinate
\footnote{Vacuum expectation values (VEV's) of 
these gauge invariant 
operators are in one to one corespondence to 
the solutions of D-flatness 
conditions \cite{LT}}. It is in general troublesome to do this task. 
However,  if the gauge symmetry breaking pattern is known 
at generic points in the moduli space, one can easily identify these 
gauge invariant operators. We illustrate below how it works 
in the present model. 
The gauge symmetry breaking pattern we utilize is \cite{Slansky}
\begin{eqnarray}
SO(12) &\stackrel{<32>}{\longrightarrow}& 
SU(6) \stackrel{<12>}{\longrightarrow} 
SU(5) \stackrel{<12>}{\longrightarrow} 
SU(4) \nonumber \\ 
&\stackrel{<12>}{\longrightarrow}&
SU(3) \stackrel{<12>}{\longrightarrow} 
SU(2) \stackrel{<12>}{\longrightarrow} 1.
\end{eqnarray}
With this information in hand, 
counting degrees of freedom of gauge invariant operators 
is nothing but a group theoretical exercise. 
We display in Table 2 parton degrees of freedom, 
unbroken subgroups, eaten degrees of freedom by Higgs mechanism 
and hadron degrees of freedom. 
%
\begin{table}[hbtp]
\begin{center}
    \begin{tabular}{ccccc}
\hline
$N_f$ & Parton & Unbroken & Eaten & Hadron \\ 
 & DOF & Subgroup & DOF & DOF \\ 
\hline
0 & 32 & SU(6) & $66-35=31$ & 1 \\ 
1 & 44 & SU(5) & $66-24=42$ & 2 \\ 
2 & 56 & SU(4) & $66-15=51$ & 5 \\ 
3 & 68 & SU(3) & $66-8=58$ & 10 \\ 
4 & 80 & SU(2) & $66-3=63$ & 17 \\ 
5 & 92 & 1 & 66 & 26 \\ 
6 & 104 & 1 & 66 & 38 \\ 
7 & 116 & 1 & 66 & 50 \\
\hline
\end{tabular}
\caption[inv]{Degrees of freedom of independent 
gauge invariants}
\end{center}
\end{table}
%
We are now in a position to construct gauge invariant 
operators explicitly. 
Before doing this, we need to notice that $SO(12)$ 
spinor product decomposes 
into the following irreducible representations 
\begin{equation}
\label{deco}
32 \times 32 = [0]_A + [2]_S + [4]_A + \tilde{[6]}_S 
\end{equation}
where [n] represents rank-n antisymmetric tensor, 
subscripts ``A" and ``S" mean antisymmetry and symmetry under 
spinor exchange, and the tilde of the last term implies that 
the rank-6 tensor is self-dual. 
Since our model has only one spinor, gauge invariant operators 
can include $[2]_S$ and $\tilde{[6]}_S$ in Eq.(\ref{deco}).

Taking this into account, we can construct gauge invariant 
composites as follows\footnote{The representations and the charge 
in Eq.(\ref{inv}) are those under Eq.(\ref{af})} 
\begin{eqnarray}
\label{inv}
L &=& \frac{1}{2!2!} 
(Q^T \Gamma^{[\mu} \Gamma^{\nu]} C Q) 
(Q^T \Gamma_{[\mu} \Gamma_{\nu]} C Q) 
\sim ({\bf 1}; {\bf 1}; 4N_f; 4R), \nonumber \\ 
M^{(ij)} &=& (V^{i \mu})^T V^j_{\mu} 
\sim ({\bf 1}; \sym; -8; 2R), \nonumber \\ 
N^{[ij]} &=& \frac{1}{2} Q^T /{\!\!\! V^i} /{\!\!\! V^j} C Q 
\sim ({\bf 1}; \antisym; 2N_f-8; 4R), \nonumber \\ 
P^{[ijklmn]} &=& \frac{1}{6!} 
Q^T /{\!\!\! V^{[i}} /{\!\!\! V^j} /{\!\!\! V^{k}} 
/{\!\!\! V^{l}} /{\!\!\! V^m} /{\!\!\! V^{n]}} C Q
\sim ({\bf 1}; \antisix; 2N_f-24; 8R), \nonumber \\ 
R^{[ijklmn]} &=& \frac{1}{6!} \epsilon^{\mu_1 \cdots \mu_{12}} 
(Q^T \Gamma_{\mu_1} \Gamma^{\nu} C Q)
(Q^T \Gamma_{\nu} \Gamma_{\mu_2} \cdots \Gamma_{\mu_6} C Q) 
V_{\mu_7}^i V_{\mu_8}^j V_{\mu_9}^k V_{\mu_{10}}^l 
V_{\mu_{11}}^m V_{\mu_{12}}^n \nonumber \\ 
&\sim& ({\bf 1}; \antisix; 4N_f-24; 10R), 
\end{eqnarray}
where the square bracket means the antisymmetrization of 
the corresponding indices and $/{\!\!\! V^i} \equiv 
V_{\mu}^i \Gamma^{\mu}$ and we use here the following 
SO(12) Gamma matrices, 
\begin{center}
\begin{tabular}{ll}
$\Gamma_1 = \sigma_2 \otimes \sigma_3 \otimes \sigma_3 
\otimes \sigma_3 \otimes \sigma_3 \otimes \sigma_3$ & 
$\Gamma_2 = - \sigma_1 \otimes \sigma_3 \otimes \sigma_3 
\otimes \sigma_3 \otimes \sigma_3 \otimes \sigma_3$ \\
$\Gamma_3 = 1 \otimes \sigma_2 \otimes \sigma_3 
\otimes \sigma_3 \otimes \sigma_3 \otimes \sigma_3$ & 
$\Gamma_4 = -1 \otimes \sigma_1 \otimes \sigma_3 \otimes \sigma_3 
\otimes \sigma_3 \otimes \sigma_3$ \\
$\Gamma_5 = 1 \otimes 1 \otimes \sigma_2 
\otimes \sigma_3 \otimes \sigma_3 \otimes \sigma_3$ & 
$\Gamma_6 = -1 \otimes 1 \otimes \sigma_1 
\otimes \sigma_3 \otimes \sigma_3 \otimes \sigma_3$ \\
$\Gamma_7 = 1 \otimes 1 \otimes 1 \otimes \sigma_2 \otimes 
\sigma_3 \otimes \sigma_3$ & 
$\Gamma_8 = -1 \otimes 1 \otimes 1 \otimes \sigma_1 
\otimes \sigma_3 \otimes \sigma_3$ \\ 
$\Gamma_9 = 1 \otimes 1 \otimes 1 \otimes 1 \otimes 
\sigma_2 \otimes \sigma_3$ & 
$\Gamma_{10} = -1 \otimes 1 \otimes 1 \otimes 1 \otimes 
\sigma_1 \otimes \sigma_3$ \\
$\Gamma_{11} = 1 \otimes 1 \otimes 1 \otimes 1 \otimes 1 \otimes 
\sigma_2$ & 
$\Gamma_{12} = -1 \otimes 1 \otimes 1 \otimes 1 \otimes 1 \otimes 
\sigma_1$ \\ 
$\Gamma_{13} = \sigma_3 \otimes \sigma_3 \otimes \sigma_3 
\otimes \sigma_3 \otimes \sigma_3 \otimes \sigma_3$ & 
\end{tabular}
\end{center}
and $C$ is a charge conjugation matrix.

In order to see that these gauge invariant operators are 
in fact the coodinates of the moduli space, one has to check 
whether the total degrees of freedom of these gauge invariant 
operators in eq (\ref{inv}) coincide with hadronic degrees of 
freedom. 
%
\begin{table}[hbtp]
\begin{center}
    \begin{tabular}{cccccccc}
\hline
$N_f$ & Hadron DOF & L & M & N & P & R & Constraints \\ 
\hline
0 & 1 & 1 & 0 & 0 & 0 & 0 & \\ 
1 & 2 & 1 & 1 & 0 & 0 & 0 & \\ 
2 & 5 & 1 & 3 & 1 & 0 & 0 & \\ 
3 & 10 & 1 & 6 & 3 & 0 & 0 & \\ 
4 & 17 & 1 & 10 & 6 & 0 & 0 & \\ 
5 & 26 & 1 & 15 & 10 & 0 & 0 & \\ 
6 & 38 & 1 & 21 & 15 & 1 & 1 & -1\\ 
7 & 50 & 1 & 28 & 21 & 7 & 7 & -14 \\
\hline
\end{tabular}
\caption[hadron]{Hadron degree of freedom count}
\end{center}
\end{table}
%
In Table 3, these degrees of freedom are listed as a function of 
$N_f$. For $N_f \le 5$, hadronic degrees of freedom and 
that of $L, M, N, P$ and $R$ agree with each other. For $N_f=6$, 
degrees of freedom of $L, M, N, P$ and $R$ 
are larger than those of hadrons by one. 
This implies that $L, M, N, P$ and $R$ are not 
independent and a single constraint among them exists. 
This statement is also consistent with the previous argument for 
dynamically generated superpotentials. 
For $N_f=7$, 14 constraints are expected to come from 
the equations of motion as in $N_f=N_c+1$ SUSY QCD.

We can obtain more non-trivial support which convinces us 
that gauge invariant operators $L, M, N , P$ and $R$ are moduli. 
One of the powerful methods to study the low energy spectrum 
is 't Hooft anomaly matching \cite{'t Hooft}. 
To see that anomalies between elementary fields and 
composite ones match, 
we take anomaly free symmetry group instead of Eq.(\ref{sym}) 
\begin{equation}
\label{af}
G_{AF} = SO(12)_{gauge} \times [SU(N_f) \times U(1) \times 
U(1)_R]_{global}
\end{equation}
where new $U(1)$ and $U(1)_R$ are linear combinations of 
the original $U(1)$'s in Eq.(\ref{sym}). 
Matter superfields transform under Eq.(\ref{af}) as 
\begin{eqnarray}
\label{matter1}
V^i_{\mu} &\sim& ({\bf 12}, \fund, -4, R), \\
Q_{\alpha} &\sim& ({\bf 32}, {\bf1}, N_f, R)
\end{eqnarray}
where $R=N_f-6/N_f+4$. We can calculate anomalies and 
see that anomalies match for $N_f=7$; $SU(7)^3: 12 A(\fund), 
SU(7)^2U(1)_R: -\frac{120}{11} \mu(\fund), SU(7)^2U(1): 
-48\mu(\fund), U(1)_R: -\frac{434}{11}, U(1): -112, 
U(1)_R^2U(1): \frac{2128}{121}, U(1)_RU(1)^2: -\frac{29120}{11}, 
U(1)_R^3: -\frac{28154}{1331}, U(1)^3: 5600$, 
where $A(\fund)$ and $\mu(\fund)$ are cubic and quadratic 
Dynkin indices for fundamental representation of $SU(N_f)$. 
Recalling that anomalies are saturated in $N_f=N_c+1$ SUSY QCD, 
this coindence for $N_f=7$ is very natural and gives a strong support 
that the theory in this case is in ``s-confinement" phase.

For $N_f \ge 8$, it is impossible to satisfy anomaly matching 
conditions without violating $N_f \le 7$ result 
even if other gauge invariant operators are added. 
This implies that a confining phase of this model terminates 
at $N_f=7$.

\section{Low energy superpotentials}

In this section, we determine explicitly low energy superpotenials 
in terms of $L, M, N, P$ and $R$. 
Since we know which gauge invariant operators are moduli 
in the previous section, it is straightforward to work out 
what should be in the superpotential. 
Following Ref. \cite{so11}, we first determine 
the quantum deformed constraint in $N_f=6$ theory. 
Dimensional analysis, symmetries and holomorphy restrict 
the superpotential as follows 
\begin{eqnarray}
\label{Nf6}
W_{N_f=6} &=& X (R^2 + P^2 L + 2P {\rm Pf}N + L^2 {\rm det}M \nonumber \\ 
&& + \frac{1}{2!4!} \epsilon_{i_1i_2i_3i_4i_5i_6} 
\epsilon_{j_1j_2j_3j_4j_5j_6} 
L N^{i_1j_1} N^{i_2j_2} M^{i_3j_3} M^{i_4j_4} M^{i_5j_5} M^{i_6j_6} 
\nonumber \\ 
&& + \frac{1}{4!2!} \epsilon_{i_1i_2i_3i_4i_5i_6} 
\epsilon_{j_1j_2j_3j_4j_5j_6} 
N^{i_1j_1} N^{i_2j_2} N^{i_3j_3} N^{i_4j_4} M^{i_5j_5} M^{i_6j_6} 
- \Lambda_6^{20} ) 
\end{eqnarray}
where $\Lambda_6$ is the strong coupling scale of $SO(12)$ 
gauge theory with 6 vector flavors and a spinor. 
Coefficients of each terms are determined so that it reproduce 
the superpotential in $SO(11)$ gauge theory with 5 flavors 
\cite{so11}.(If VEV $<V^6_{12}>\ne 0$ is given, $SO(12)$ gauge 
theory with 6 flavors under consideration here reduces to 
$SO(11)$ gauge theory with 5 flavors)

One may also determine these coefficients by using 
the symmetry breaking along the spinor flat direction 
$SO(12) \stackrel{<32>}{\longrightarrow} SU(6), 
{\rm explicitly} <{\bf 32}>^T = (0,a,0,\cdots,0,a,0)$. 
$V^i_{\mu}$ decomposes into ${\bf 6}+{\bf \bar{6}}$ 
under $SU(6)$, which is explicitly as 
\begin{equation}
V_{\mu} = \left(
\begin{array}{c}
q_1+\bar{q}_1 \\
i(q_1-\bar{q}_1) \\
q_2+\bar{q}_2 \\
i(q_2-\bar{q}_2) \\
q_3+\bar{q}_3 \\
i(q_3-\bar{q}_3) \\
q_4+\bar{q}_4 \\
i(q_4-\bar{q}_4) \\
q_5+\bar{q}_5 \\
i(q_5-\bar{q}_5) \\
q_6+\bar{q}_6 \\
i(q_6-\bar{q}_6) 
\end{array}
\right)
\end{equation}
where $q_i,\bar{q}_i(i=1,\cdots,6)$ mean $SU(6)$ 
quarks, antiquarks, respectively. 
The reduced theory is $SU(6)$ gauge theory with $6(\fund+\antifund)$, 
therefore it has a single quantum constraint \cite{Seiberg}.

According to this decompostion rule, $SO(12)$ gauge invariant 
operators are decomposed into the following $SU(6)$ meson $m^{ij}$, 
baryon $b$ and anti-baryon $\bar{b}$; 
\begin{eqnarray}
L &\to& 12a^4 \nonumber \\ 
M^{ij} &\to& 2(m^{ij} + m^{ji}) \nonumber \\ 
N^{[ij]} &\to& -4ia^2 (m^{ij} - m^{ji}) \nonumber \\
P &\to& 64ia^2 (b + \bar{b}) + 2ia^2 
\epsilon_{ijklmn} m^{ij}m^{kl}m^{mn} \nonumber \\
R &\to& 64a^4 (b - \bar{b}) 
\end{eqnarray}
Using this information, one can also determine coefficients 
so that ${\rm det}m - b\bar{b} = \Lambda^{12}$ be reproduced.

The superpotential of $N_f=7$ case can be found in a similar way. 
In this case, the superpotential must have the following features.
1. It is smooth everywhere on the moduli space. 
2. Equations of motion give classical constraints 
among vectors and a spinor. 3. Adding mass term for one vector 
flavor to this superpotential and integrating out 
this massive vector, 
the superpotential (\ref{Nf6}) must be reproduced. 
The result is\footnote{Although this superpotential has 
already been derived 
in Ref. \cite{s-conf} by using the index argument, 
a $PMN^3$ term was missing. 
Without this term, the result of Ref. \cite{so11} cannot be 
correctly recovered.}
\begin{eqnarray}
\label{Nf7}
W_{N_f=7} &=& \frac{1}{\Lambda^{19}} (M^{ij} R_i R_j 
-2i N^{ij} P_i R_j + 
L P_i P_j M^{ij} \nonumber \\ 
&& + L^2 {\rm det}M + 
\frac{1}{3!2^2} \epsilon_{i_1 \cdots i_7} P_j M^{ji_1} 
N^{i_2i_3}N^{i_4i_5}N^{i_6i_7} \nonumber \\
&& + \frac{1}{3!4!} 
\epsilon_{i_1i_2i_3i_4i_5i_6i_7} \epsilon_{j_1j_2j_3j_4j_5j_6j_7} 
L N^{i_1j_1} N^{i_2j_2} 
M^{i_3j_3} M^{i_4j_4} M^{i_5j_5} M^{i_6j_6} M^{i_7j_7} 
\nonumber \\ 
&& + \frac{1}{4!3!} 
\epsilon_{i_1i_2i_3i_4i_5i_6i_7} \epsilon_{j_1j_2j_3j_4j_5j_6j_7} 
N^{i_1j_1} N^{i_2j_2} N^{i_3j_3} N^{i_4j_4} 
M^{i_5j_5} M^{i_6j_6} M^{i_7j_7} ).
\end{eqnarray}

By adding the mass terms for vector fields 
$\delta W = m_{ij}M^{ij}$ 
to the superpotential (\ref{Nf6}) and integrating out 
each massive vectors successively, 
we can readily derive the superpotentials 
for $N_f \le 5$ systematically. 
As a matter of fact, we obtain 
\begin{eqnarray}
\label{Nfle5}
W_{N_f=5} &=& \Lambda_5^{21}/( L^2 {\rm det}M + 
\frac{1}{2!3!} L N^{i_1j_1} N^{i_2j_2} 
M^{i_3j_3} M^{i_4j_4} M^{i_5j_5} 
\epsilon_{i_1i_2i_3i_4i_5} 
\epsilon_{j_1j_2j_3j_4j_5} \nonumber \\ 
&& + \frac{1}{4!} N^{i_1j_1} N^{i_2j_2} 
N^{i_3j_3} N^{i_4j_4} M^{i_5j_5} 
\epsilon_{i_1i_2i_3i_4i_5} 
\epsilon_{j_1j_2j_3j_4j_5} ), \nonumber \\
W_{N_f=4} &=& 2 \left( \frac{\Lambda_4^{22}}{L^2 
{\rm det}M + 
\frac{1}{2!2!} L N^{i_1j_1} N^{i_2j_2} 
M^{i_3j_3} M^{i_4j_4} 
\epsilon_{i_1i_2i_3i_4} \epsilon_{j_1j_2j_3j_4} 
+ ({\rm Pf}N)^2} 
\right)^{1/2}, \nonumber \\
W_{N_f=3} &=& 3 \left( \frac{\Lambda_3^{23}}{L^2 {\rm det}M + 
\frac{1}{2!} L M^{i_1j_1} N^{i_2j_2} N^{i_3j_3} 
\epsilon_{i_1i_2i_3} \epsilon_{j_1j_2j_3}} 
\right)^{1/3}, \nonumber \\
W_{N_f=2} &=& 4 \left( \frac{\Lambda_2^{24}}{L^2 {\rm det}M + 
L N^2} \right)^{1/4}, \nonumber \\
W_{N_f=1} &=& 5 \left( \frac{\Lambda_1^{25}}{L^2 M} 
\right)^{1/5}, \nonumber \\
W_{N_f=0} &=& 6 \left( \frac{\Lambda_0^{26}}{L^2} 
\right)^{1/6}, 
\end{eqnarray}
where the strong coupling scales for each flavor are related to 
each other through one-loop matching of gauge coupling as follows 
\begin{eqnarray}
\Lambda_0^{26} &=& m_{11} \Lambda_1^{25} = m_{11}m_{22} \Lambda_2^{24} = 
m_{11}m_{22}m_{33} \Lambda_3^{23} \nonumber \\ 
&=& m_{11}m_{22}m_{33}m_{44} \Lambda_4^{22} = 
m_{11}m_{22}m_{33}m_{44}m_{55} \Lambda_5^{21} \nonumber \\ 
&=& m_{11}m_{22}m_{33}m_{44}m_{55}m_{66} \Lambda_6^{20} = 
m_{11}m_{22}m_{33}m_{44}m_{55}m_{66}m_{77} \Lambda_7^{19}. 
\end{eqnarray}
It is worth to note that one can confirm the above superpotentials 
(\ref{Nfle5}) 
to recover correctly the superpotentials for $N_f \le 4$ 
in $SO(11)$ theory \cite{so11}, which is obtained 
when one flavor vector field has non-vanishing VEV.

Before closing this section, we briefly discuss dynamical 
supersymmetry breaking. For $N_f=6$, if we take the tree 
level superpotential as 
\begin{equation}
W_{tree} = \lambda_1 S_1 V^2 + \lambda_2 S_2 Q^2 V^6 
+ \lambda_3 S_3 Q^4 V^6
\end{equation}
where $S_{1,2,3}$ are singlet superfields, 
then equations of motion with respect to $S_{1,2,3}$ and 
the quantum constraint (\ref{Nf6}) are incompatible. 
Therefore, supersymmetry is dynamically broken \cite{IYIT}. 
For $N_f=0$, since we cannot add terms which lift 
a classical flat direction ($i.e. L$) 
preserving $U(1)_R$ to the tree level superpotential, 
supersymmetry remains unbroken. 
The same argument seems to be applicable for $1 \le N_f \le 5$
\footnote{In Ref. \cite{s-conf}, the authors construct a model with 
dynamical supersymmetry breaking for $N_f=1$ by promoting a 
global $U(1)$ to a local $U(1)$ and adding singlets to cancel 
$U(1)$ gauge anomaly.}.

\section{Summary}
%
In this paper, we have studied the confining phase 
in ${\cal N}=1$ $SO(12)$ SUSY gauge theory with 
$N_f \le 7$ vectors and a spinor. 
Utilizing the gauge symmetry breaking pattern at generic 
points on the moduli space which plays a crucial role 
in our study, we have identified gauge 
invariant operators which behave as the moduli coordinate. 
Then we have derived explicitly low energy superpotentials 
for $N_f\le7$.

Some clues in search for duals are obtained from the 
results of this work. Let us suppose that the dual with 
the gauge group $\tilde{G}$ exists for $N_f \ge 8$. 
The original $SO(12)$ theory breaks down to 
$SU(6)+N_f(\fund+\antifund)$ along the spinor flat direction. 
On the other hand, the gauge group of the dual theory is 
usually unbroken since in the dual theory the gauge invariant 
operators develop VEV. 
Since $SU(6)+N_f(\fund+\antifund)(N_f \ge 8)$ theory is dual to 
$SU(N_f-6)+N_f(\fund+\antifund)$ \cite{Seiberg1}, 
we can guess that at least 
$\tilde{G}$ must include $SU(N_f-6)$ as a subgroup to preserve 
the duality along this direction. 
Furthermore, the superpotential in the dual theory must recover 
$N_f=7$ superpotential. We also note that $N_f=8$ case 
in our model is known to be self-dual \cite{selfdual}.

Although it seems to be quite difficult to find a dual 
which is compatible with the above requirements, 
we hope that this work will provide useful 
informations to search for 
the dual to $SO(N_c)(N_c\ge11)$ theory.

\begin{center}
{\bf Acknowlegdements}
\end{center}
The author thanks S. Kitakado and T. Matsuoka for careful 
reading of the manuscript and for useful discussions.

\end{document}